\documentclass{svjour3}                     
\smartqed  
\usepackage{graphicx}
\usepackage{url}
\usepackage{color}
\definecolor{red}{rgb}{1.0,0.0,0.0}
\begin{document}

\title{Comprehensive Analysis of Market Conditions in the Foreign
Exchange Market}
\subtitle{Fluctuation Scaling and Variance-Covariance Matrix}

\titlerunning{Comprehensive Analysis on Market Conditions ...}

\author{Aki-Hiro Sato, Takaki Hayashi, and Janusz A. Ho\l yst}

\authorrunning{A.-H. Sato, T. Hayashi, and J.A. Ho\l yst} 
\institute{A.-H. Sato \at
              Graduate School of Informatics, Kyoto University,
              Yoshida Honmachi, Sakyo-ku, Kyoto, 606-8501, JAPAN \\
              Tel.: +81-75-753-5515\\
              Fax: +81-75-753-4919\\
              \email{aki@i.kyoto-u.ac.jp}             \\
	      T. Hayashi \at
              Graduate School of Business Administration, Keio University \\
              4-1-1 Hiyoshi, Yokohama, Kanagawa 223-8526, JAPAN \\
              \email{takaki@kbs.keio.ac.jp} \\
              J.A. Ho\l yst \at
              Faculty of Physics and Center of Excellence for Complex
              Systems Research, \\ 
	      Warsaw University of Technology, Koszykowa 75, PL-00-662,
              Warsaw, POLAND \\
              \email{jholyst@if.pw.edu.pl}           
}

\date{Received: 24 September 2010 / Accepted: 3 March 2012}
\maketitle

\begin{abstract}
We investigate quotation and transaction activities 
in the foreign exchange market for every week during the period of
June 2007 to December 2010. A scaling relationship between the mean
values of number of quotations (or
number of transactions) for various currency pairs
and the corresponding standard deviations holds for a majority of the
weeks. However, the scaling 
breaks in some time intervals, which is related to the emergence of
market shocks. There is a monotonous relationship between values of
scaling indices and global averages of currency pair cross-correlations 
when both quantities are observed for various
window lengths $\Delta t$.   
\end{abstract}

\section{Introduction}
\label{sec:introduction}
Recent development of information and communication technology
empowers us both to collect and analyse a large amount of
data on socio-economic systems. Several researchers refer to the present
epoch driven by this technological changes  as {\it the information
explosion era}~\cite{info-plosion}.

This progress enables us to monitor many details of financial assets
and to record the information as a high frequency multivariate time series. 

Recently, investigations of foreign exchange rates have been
conducted by numerous researchers using various approaches based on
statistical physics and time series analysis
~\cite{Drozdz:2010,Rebitzky:10,Kaltwasser:10,Alfrano:06,Gworek:09,Liu:2010,Sazuka:09,Inoue:10,Ohnishi:04,Hashimoto:10}. Dro\.zd\.z
et al. showed that exchange rate return fluctuations for main currency
pairs are well described by non-extensive statistics and possess
multifractal characteristics~\cite{Drozdz:2010}. Rebitzky studied
the influence of macroeconomic news on exchange
rates~\cite{Rebitzky:10}. He concluded the following three points: (1) fundamental
news do matter, (2) news influences exchange rates by
incorporating common information into prices directly and indirectly
based upon order flow and (3) the impact of fundamental news on exchange
rates is fairly 
stable over time. Kaltwasser~\cite{Kaltwasser:10} estimated the herding
tendency in the foreign exchange market for three currency pairs using
the extended Alfrano-Lux model~\cite{Alfrano:06} and computed several
unconditional moments of corresponding daily log-returns. Gworek et
al. analysed the exchange rate returns of 38 currencies (including
gold)~\cite{Gworek:09}. They examined the cross-correlations between
the returns of various currency pairs as well as between their signs, and 
in this way, they constructed a corresponding Minimal Spanning
Tree for several base currencies. Liu et al. applied cross-sample
entropy (Cross-SampEn) to compare two different time series in order to asses their degree of
asynchrony to the daily log-return time series of foreign exchange rates
in the currency markets~\cite{Liu:2010}. They compared the correlations
between time series with Cross-SampEn and showed that Cross-SampEn is
superior in describing the correlations.

Moreover, theoretical models~\cite{Sazuka:09,Inoue:10}
and empirical analysis~\cite{Ohnishi:04,Hashimoto:10} on the tick-by-tick
data of the foreign exchange market have been considered. Ohnishi et  
al.~\cite{Ohnishi:04} proposed a weighted-moving-average analysis for
the tick-by-tick data of yen-dollar exchange rates. They concluded that
the weights decay exponentially with a time scale less than 2 min, implying
that dealers are watching only the very recent market state. Hashimoto and
Ito examined market impact of Japanese macroeconomic statistic news
within minutes of their announcements on the yen-dollar exchange
rates~\cite{Hashimoto:10}. They found clear increases in the number 
of deals and price volatility immediately after such announcements.

In principle, information arrivals can be never observed comprehensively;
however, one can register quotations(transactions) from market
participants. Perceived information determines investor attitude, but
the impact of news on trading decisions is an open
question~\cite{Kyle:85}.

Nevertheless, we believe that it is important to confirm
whether correlations exist between the number of quotations
and transactions. In this study, we focus on the number of quotations
or transactions for several currency pairs quoted in the foreign
exchange market and we treat them as a multivariate time
series that characterise system dynamics. Bonanno et
al. investigated the spectral density of both the logarithm of prices
and the daily number of trades of a set of stocks traded in the New York Stock
Exchange~\cite{Bonanno:00}. They confirmed the $1/f^2$ behaviour of
the spectral density of stocks log-returns and detected a
$1/f$-like behaviour of the spectral density of daily trade
numbers. This shows that the number of transactions is generated by a
long memory process, similar to the volatility dynamics.  

More conceptually, the mixture distribution hypothesis is one of the most
famous explanatory models for price fluctuations in the financial 
market~\cite{Mandelbrot,Clark,Tauchen-Pitts,Richardson-Smith,Karpoff,Ane-Geman}.
Specifically, there is a positive simultaneous correlation between
transaction volume, or the number of transactions, and
returns~\cite{Mandelbrot,Karpoff,Ane-Geman}. Furthermore, 
Mandelbrot-Taylor~\cite{Mandelbrot} and Clark~\cite{Clark} proposed  
the concept of time changes or subordinated processes in order to explain
the fat-tailedness of returns. It was reported that the
number of trades is profoundly related to volatility. According to
the normal variance mixture models in finance~\cite{Fergusson:2006},
despite of the heteroskedacity of volatility with long memory,
unconditional distribution of log-returns can be fitted to the mixture
of normal distributions with an unconditional distribution of
volatility. The same theory has been recently proposed in statistical physics
by Beck to investigate superstatistics~\cite{Beck}. Therefore, it
is important to capture the cross-sectional structure of quotations and
transactions to understand their volatility.

In studies on financial time series, several researchers have taken the
stance that traded volume, or the number of transactions (quotations), is
a proxy variable for unobservable information arrivals on market
participants~\cite{Clark,Tauchen-Pitts,Karpoff}. 

Researchers of econophysics focus on several types of scaling
relationships observed in the financial markets in 
order to understand the behaviour of market
participants~\cite{Micciche:02,Podobnik:00,Wang:09,Eisler:06}. Specifically,
a scaling relationship between mean values of
money flow and their standard deviation
(``Taylor's power law'' or ``fluctuation scaling'') 
is a useful scaling relationship when conducting
multi-dimensional analysis on financial markets.  

A scaling relationship between a mean of constituents'
flows at the $i$-th observation point $\mu_i$ and their
standard deviation $\sigma_i$ is referred to as
fluctuation scaling:  
\begin{equation}
\mu_i = A \sigma_i^{\alpha}, 
\label{eq:scaling}
\end{equation}
where $A$ represents a positive constant, and $\alpha$ is a scaling exponent
taking a value ranging from 1/2 to 1. The pioneering work
on fluctuation scaling is known as a Taylor's study in
ecology~\cite{Taylor}. After his study, the scaling relationship between
means and standard deviations has been found in a wide spectrum of
fields. Recently, Menezes and Bar\'abasi have reported the scaling
relationship on traffic flows, river flows and browsing activities of
Internet~\cite{Menezes}.  Eisler and Kert\'esz
investigated fluctuation scaling of the traded value and dependence of
the scaling exponent on a time scale~\cite{Eisler:06}. More recently
they computed the scaling relationship of the traded volume for 2647
stocks listed in the New York Stock Exchange (NYSE)~\cite{Eisler:08} and
4039 listed in the NASDAQ in the TAQ database, which records all
transactions of the NYSE and NASDAQ during the period from 2000 to
2002. They found a clear scaling relationship at these stock
markets. They clarified that the Hurst exponents and the scaling
exponents depend on the window length. Eisler et al. investigated
fluctuation scaling on the volumes traded in the stock exchange market.

Furthermore, some researchers have examined cross-correlations among various 
quantities of the financial
markets~\cite{Mantegna:99,Podnik:09,Horvatic:09,Duan:11}.  
Podobnik et al. investigated volume growth rate and volume changes for
14,981 daily records of the Standard and Poor's (S\&P) 500 index over
a 59-year period (1950-2009)~\cite{Podnik:09}. Using detrended
cross-correlation analysis, they found that there are
power-law cross-correlations between them. Bonanno et al. studied
correlation-based network analysis of financial
equities~\cite{Bonanno:04}.

However, it is not obvious whether a stable scaling
relationship exists in the foreign exchange market. This is the first
focus of our paper. Following the Taylor methodology, we examine the scaling
relationship between the mean of the numbers of quotations and
transactions and their variance.

We suppose that there should exist a relationship between
fluctuation scaling for the market activity and cross-correlations at 
the foreign exchange market. This is the second focus of this article.

This article is organised as follows. In
Sec. \ref{sec:empirical-analysis}, we analyse the scaling 
exponents and cross-correlation matrices for a selected period. In
Sec. \ref{sec:long-term}, we conduct a long-term analysis
of both the scaling exponents and the global average of the
cross-correlations of both quotation and transaction numbers.
Sec. \ref{sec:conclusion} is devoted to concluding remarks and used to
address future works.

\section{The scaling exponent and global average of correlations}
\label{sec:empirical-analysis}
In this section, we focus on the number of quotations
arriving at the matching engine of the EBS Platform and
the number of transactions occurring there per unit
time. We use comprehensive data (ICAP EBS Data Mine 
Level 1.0) collected by the EBS Platform, one of the most popular
electronic brokerage systems, which is exploited by  over 2,000 traders
at about 800 dealing rooms across the globe~\cite{ICAP}. The data include
records for orders (BID/OFFER) and transactions of currencies and
precious metals during a period from June 2007 to December 2010 (43
months)~\footnote{We found 43 kinds of  
currencies and 68 currency pairs during the whole
observation period: AUD/NZD, AUD/USD, CHF/JPY, EUR/CHF,
EUR/CZK, EUR/DKK, EUR/GBP, EUR/HUF,  
EUR/ISK, EUR/JPY, EUR/NOK, EUR/PLN, EUR/SEK, EUR/SKK, EUR/USD, EUR/ZAR,
GBP/USD, NZD/USD, USD/CAD, USD/CHF, USD/HKD, USD/JPY, USD/MXC, USD/MXN,
USD/MXT, USD/PLN, USD/RUB, USD/SGD, USD/ZAR, XAG/USD, XAU/USD, XPD/USD, 
XPT/USD, GBP/JPY, GBP/CHF, USD/TRY, XAU/JPY, AUD/JPY, USD/THB, CAD/JPY,
NZD/JPY, EUR/RUB, ZAR/JPY, EUR/AUD, EUR/CAD, EUR/RON, GBP/AUD, USD/SEK,
BKT/RUB, USD/DKK, USD/NOK, USD/ILS, SAU/USD, DLR/KES, DLR/KET, USD/AED,
USD/BHD, USD/KWD, USD/SAR, EUQ/CHF, EUQ/JPY, USQ/CHF, EUQ/USD, USQ/JPY,
USD/CNH, CNH/HKD, CNH/JPY and EUR/CNH. The total number of both
quotations and transaction recorded during this observation duration is
318,871,428.}. During this period, the global financial system suffered
from the following significant macroeconomic shocks and crises: (I)
Paribas shock (Aug. 2007), (II) Bear Stearns shock (Feb. 2008), (III)
sub-prime crisis driven by Lehman shock (Sep. 2008 to Mar. 2009) and
(IV) Euro crisis (Apr. to May 2010).

Let $P_{j,\Delta t}(k)$ and $D_{j,\Delta t}(k)$ be the
number of quotations and transactions for exchange between the pairwise
currencies $j$ during an observation period between $k\Delta t$ and
$(k+1)\Delta t$ $(k=0,1,\ldots,Q-1)$, respectively. Let us further 
assume that $P_{j,\Delta t}(k)$ and $D_{j,\Delta t}$ are locally
stationary variables in the time interval $[0,(Q-1)\Delta t]$. A
temporal mean and a variance-covariance matrix are expressed by the
notation $\langle \cdots \rangle$ and $\mbox{Cov}(\cdot,\cdot)(\tau)$:
\begin{eqnarray}
\langle X_{j,\Delta t} \rangle &=&
 \frac{1}{Q}\sum_{k=0}^{Q-1}X_{j,\Delta t}(k),
\label{eq:mean-sd}
\\
\nonumber
\mbox{Cov}\bigl(X_{i,\Delta t}, Y_{j,\Delta t}\bigr)(\tau) &=&
\left\{
\begin{array}{lll}
\frac{1}{Q+\tau}\sum_{k=0}^{Q+\tau-1}\Bigl(X_{i,\Delta t}(k-\tau) - \langle
X_{i,\Delta t} \rangle\Bigr)\Bigl(Y_{j,\Delta t}(k) - \langle Y_{j,\Delta t} \rangle
\Bigr) & (\tau < 0)\\
\frac{1}{Q-\tau}\sum_{k=0}^{Q-\tau-1}\Bigl(X_{i,\Delta t}(k) - \langle
X_{i,\Delta t} \rangle\Bigr)\Bigl(Y_{j,\Delta t}(k+\tau) - \langle Y_{j,\Delta t} \rangle
\Bigr) & (\tau \geq 0)\\
\end{array}
\right.,\\
\label{eq:cov-spd}
\end{eqnarray}
where $X$ and $Y$ are selected from the set consisting of $P$
and $D$.

Elements of the correlation matrix of both quotation or
transaction activities for various currency pairs $i$ and $j$ are, respectively,
defined in the standard way:
\begin{eqnarray}
C_{ij,\Delta t}^{(XX)}(\tau) &=&
 \frac{\mbox{Cov}\bigl(X_{i,\Delta t}, X_{j,\Delta t}\bigr)(\tau)}
{\sqrt{\mbox{Cov}\bigl(X_{i,\Delta t}, X_{i,\Delta t}\bigr)(\tau)
\mbox{Cov}\bigl(X_{j,\Delta t}, X_{j,\Delta t}\bigr)(\tau)}}.
\label{eq:correlation}
\end{eqnarray}
The mean value of its nondiagonal components describes the global
average of simultaneous cross-correlation coefficients:
\begin{equation}
\langle C_{\Delta t}^{(X)}\rangle = \frac{2}{N(N-1)}\sum_{i=1}^{N-1}
\sum_{j=i+1}^N C_{ij,\Delta t}^{(XX)}(0),
\label{eq:averaged-correlation}
\end{equation}
where $X=P$ for quotations and $X=D$ for transactions.

We shall look for a scaling relationship in the form:
\begin{equation}
\left\{
\begin{array}{lcl}
\mbox{Cov}\bigl(P_{i,\Delta t}, P_{i,\Delta t}\bigr)(0) &=& A_P
\langle P_{i,\Delta t} \rangle^{2\alpha_P} \\
\mbox{Cov}\bigl(D_{i,\Delta t}, D_{i,\Delta t}\bigr)(0) &=& A_D
\langle D_{i,\Delta t} \rangle^{2\alpha_D}
\end{array}
\right.,
\label{eq:scaling2}
\end{equation}
where $A_P$ and $A_D$ are positive constants and
$\alpha_P$ and $\alpha_D$ are scaling
exponents ranging from $1/2$ to $1$. This scaling law means that
currency pairs with a low (high) mean activity show small (large)
activity fluctuations. Values of scaling exponents $\alpha_X$ can be
estimated by using the least squared method applied to the equation:  
\begin{equation}
\log \mbox{Cov}_{XX}\bigl(X_{i,\Delta t}, X_{i,\Delta t}\bigr)(0) = \log A_X +
 2\alpha_X \log \langle X_{i,\Delta t} \rangle.
\end{equation}

\begin{figure}[h]
\centering
\includegraphics[scale=0.8]{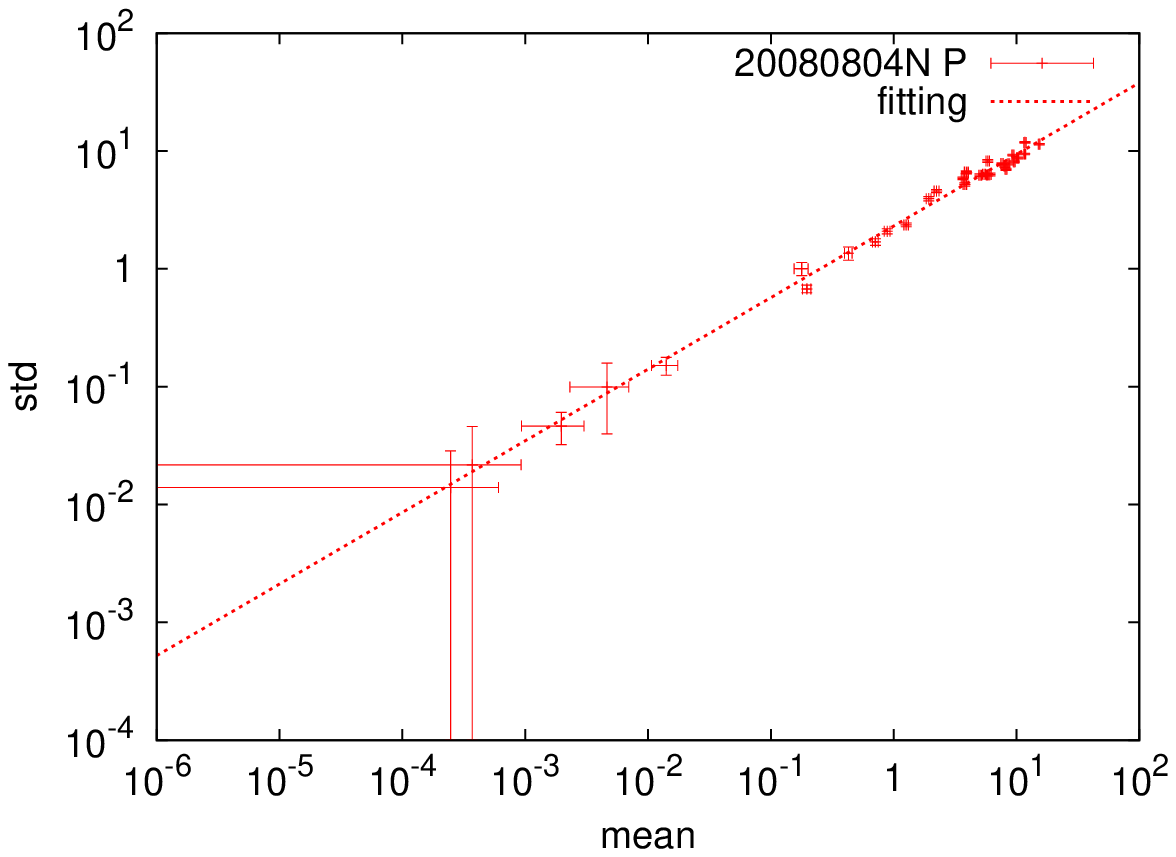}(P)
\includegraphics[scale=0.8]{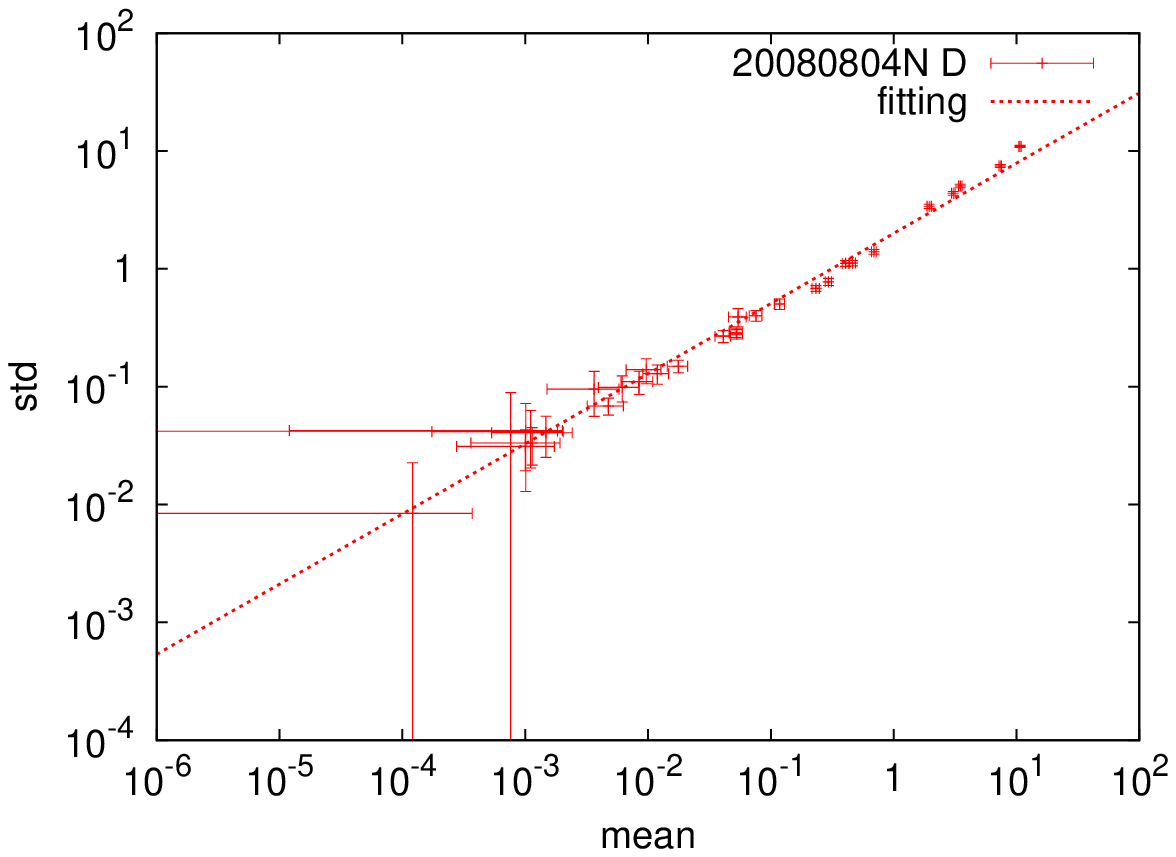}(D)
\caption{The double logarithmic plots mean of the number of quotations
 (P) and transactions (D) and their standard deviation. We use
high resolution data from 8:38, 3 August 2008, to
 21:56, 8 August 2008 (UTC). The error bars show 2-bootstrap standard
 deviations.} 
\label{fig:emp-FS}
\end{figure}

Figure \ref{fig:emp-FS} shows double logarithmic plots for standard
deviations of quotations or transactions numbers against their mean
values for investigated currency pairs during a selected period. The
error bars represent 2-bootstrap standard deviations calculated from
1,000 bootstrap samples~\cite{Efron:79}. In Ref \cite{Sato:10}, two of
the authors (A.-H. Sato and J.A. Ho\l yst) investigated the scaling
relationship with a different database of the foreign exchange market
and found fluctuation scaling for quotation activities. Similarly,
one can observe that the scaling Eq. (\ref{eq:scaling}) is well
fulfilled with the ICAP database.

\begin{figure}[bht]
\centering
\includegraphics[scale=0.82]{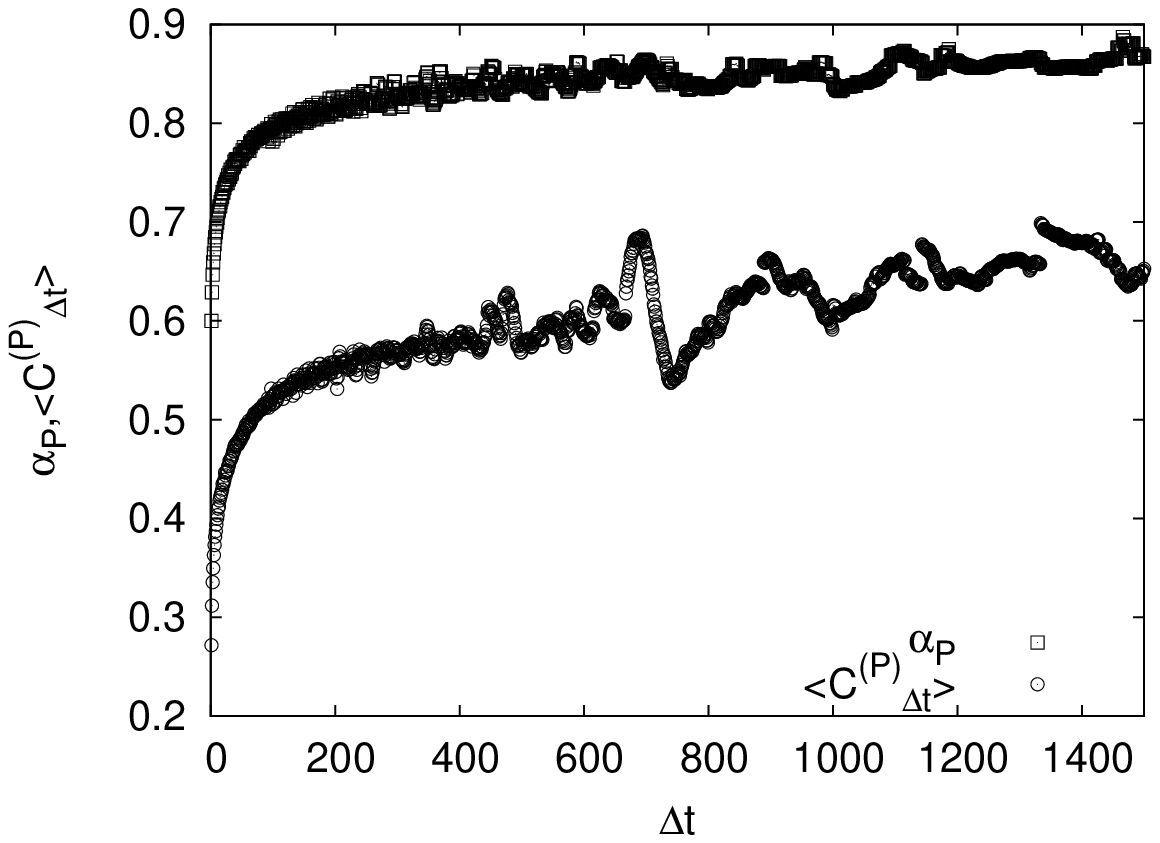}(a)
\includegraphics[scale=0.82]{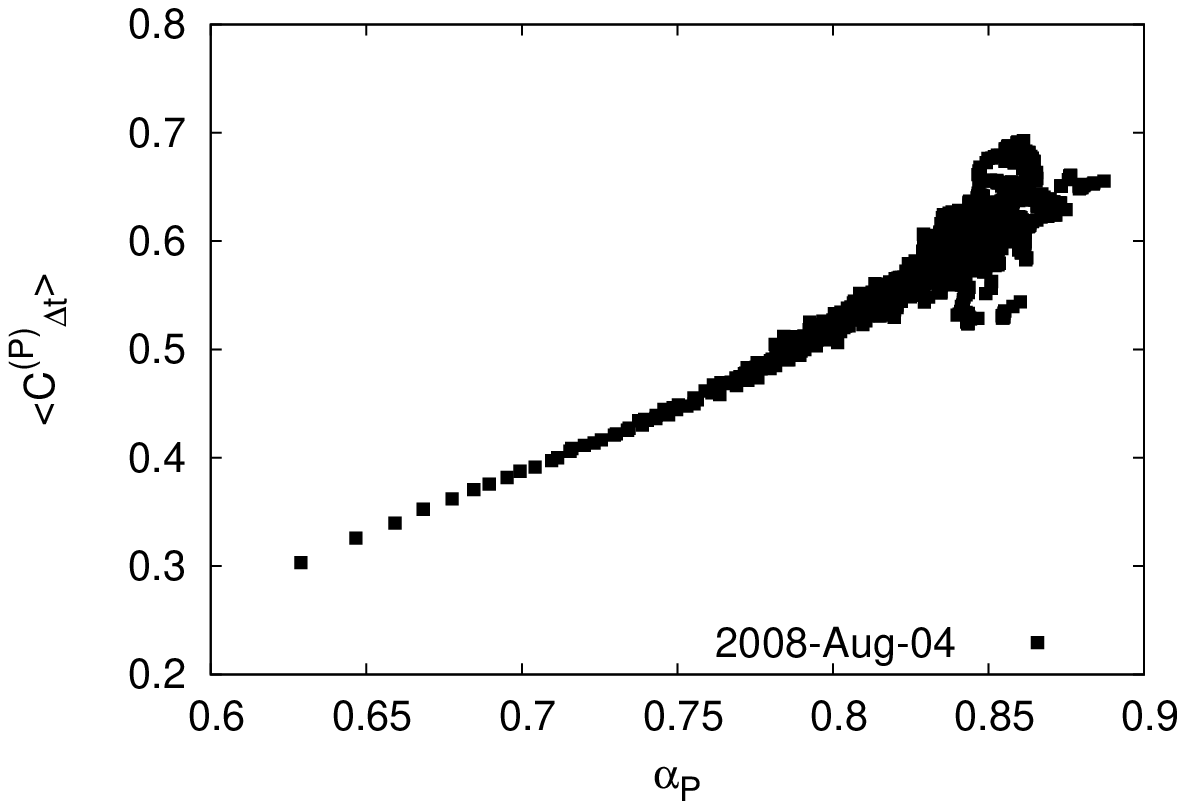}(b)
\caption{(a) The dependence of $\alpha_P$ and $\langle C_{\Delta t}^{(P)}
 \rangle$ on the values of $\Delta t$. (b) Scatter plots
 of the scaling exponents $\alpha(\Delta t)$ and the averaged
 correlation $\langle 
 C_{\Delta t}^{(P)} \rangle$. The scaling exponents $\alpha_P$ and the
 averaged correlation $\langle C_{\Delta t}^{(P)} \rangle$ are computed
 from the quotation activities during a period from 8:38, 3 August 2008, 
 to 21:56, 8 August 2008 (UTC) for the window length $\Delta t \quad
 (1 \mbox{[min]} \leq \Delta t \leq 1500 \mbox{[min]})$.}
\label{fig:dt-dependence}
\end{figure}

Figure \ref{fig:dt-dependence} shows the dependence of $\alpha_P$ and
$\langle C_{\Delta t}^{(P)} \rangle$ on the values of $\Delta t$ and a
scatter plots of $\alpha_P$ and $\langle C_{\Delta t}^{(P)} \rangle$ at
different values of $\Delta t$. Both $\alpha_P$ and
$\langle C_{\Delta t}^{(P)} \rangle$ increase as $\Delta t$ increases.  


Moreover, we extend the global average of
cross-correlation coefficients between the number of quotations and
transactions. By using Eq. (\ref{eq:cov-spd}), we define
the cross-correlation function matrix as
\begin{equation}
PD_{ij,\Delta t}(\tau) =
\frac{\mbox{Cov}\bigl(P_{i,\Delta t}, D_{j,\Delta
t}\bigr)(\tau)}{\sqrt{
\mbox{Cov}\bigl(P_{i,\Delta t}, D_{i,\Delta t}\bigr)(\tau)
\mbox{Cov}\bigl(P_{j,\Delta t}, D_{j,\Delta t}\bigr)(\tau)
}},
\label{eq:average-pd}
\end{equation}
and the global average of cross-correlations as
\begin{equation}
\langle PD_{\Delta t}\rangle(\tau) = \frac{1}{N^2}\sum_{i=1}^N\sum_{j=1}^N
PD_{ij,\Delta t}(\tau).
\end{equation}
Obviously $PD_{ij,\Delta t}(\tau)$ quantifies the anticipatory behaviour of
the quotation activities against transaction
activities.

\begin{figure}[h]
\centering
\includegraphics[scale=0.8]{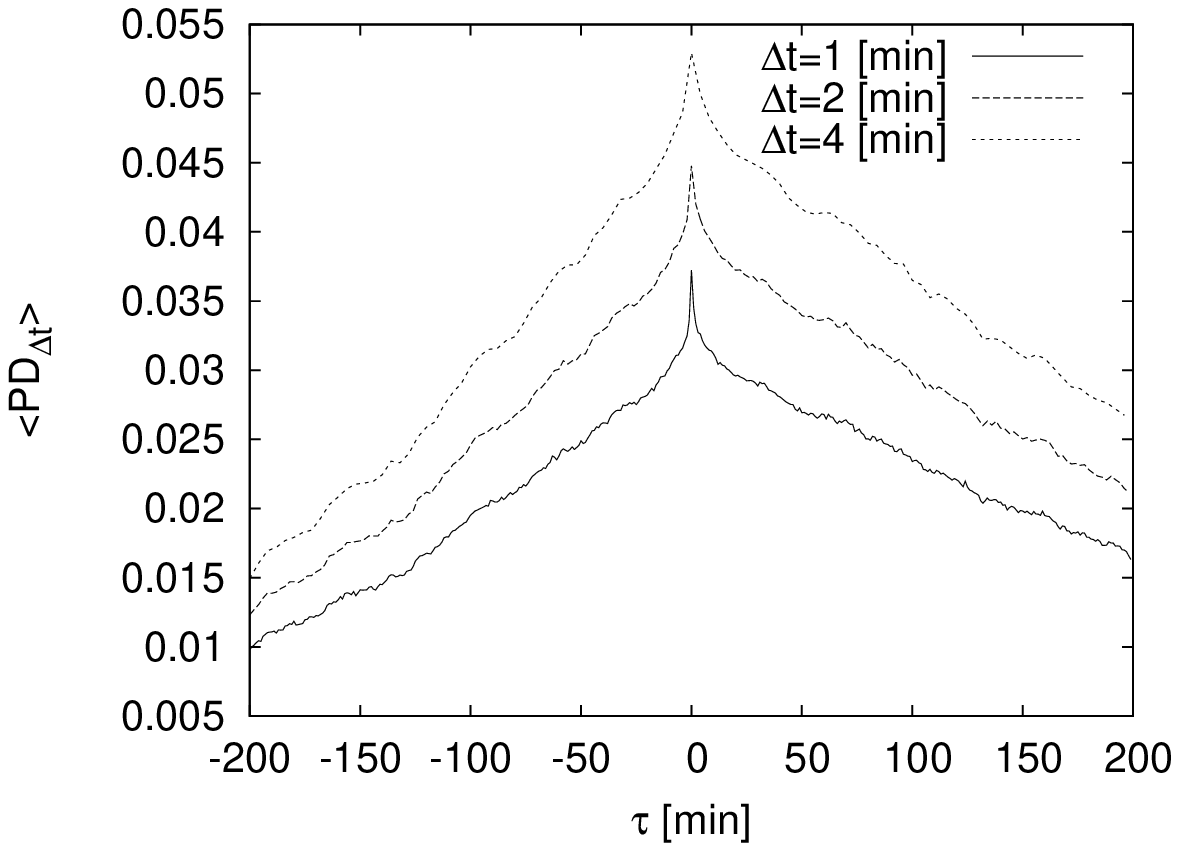}
\caption{The global average of cross-correlations
 between the quotation activities and transaction activities for
 different time windows. We use the high resolution data during a period
 from 8:38, 3 August 2008, to 21:56, 8 August 2008 (UTC).}
\label{fig:PD}
\end{figure}

Figure \ref{fig:PD} shows the global average of the cross-correlations
between the quotation activities and transaction activities. One can see
that the global average of the cross-correlations possesses a maximum
for the time difference $\tau=0$. Further, the positive
tail of the cross-correlation function decays slower in time than does
the negative one.

\section{Long-term analysis}
\label{sec:long-term}
Now we show results for longer time periods. Using ICAP EBS Data Mine
Level 1.0, we performed the investigations of the scaling exponent and
the global average of cross-correlations for both quotation and transaction
activities. Throughout this analysis, we fix $T=Q\Delta t = 10,080$
[min] and $\Delta t = 1$[min]. The empirical results are obtained from
Eqs. (\ref{eq:mean-sd}) and (\ref{eq:cov-spd}). 

We compute both the bootstrap mean of the scaling indices and their
variance for each week. We generated 1,000 bootstrap samples with 100
points from $10,080$ observation points.  
We computed the temporal variation of scaling exponents $\alpha$ for both
quotations and transactions during all the observation weeks. 
It is found that the bootstrap standard deviation of the scaling indices
is less than 0.01. They vary in range from $0.5$ to $0.7$.

In order to measure an error between empirical results
and a fitting function with the least squars method,
we define a norm of residual,
\begin{equation}
normr_{X} = \sqrt{\sum_{i=1}^N \Bigl( \log \mbox{Cov}\bigl(X_{i,\Delta t}, X_{i,\Delta
 t}\bigr)(0) - \log \hat{A}_X - 2 \hat{\alpha}_X \log \langle
 X_{i,\Delta t } \rangle \Bigr)^2},
\end{equation}
where $\hat{A}_X$ and $\hat{\alpha}_X$ ($X=\{P,D\}$) are parameter
estimates computed by using the least squared method.

\begin{figure}[phtb]
\centering
\includegraphics[scale=0.8]{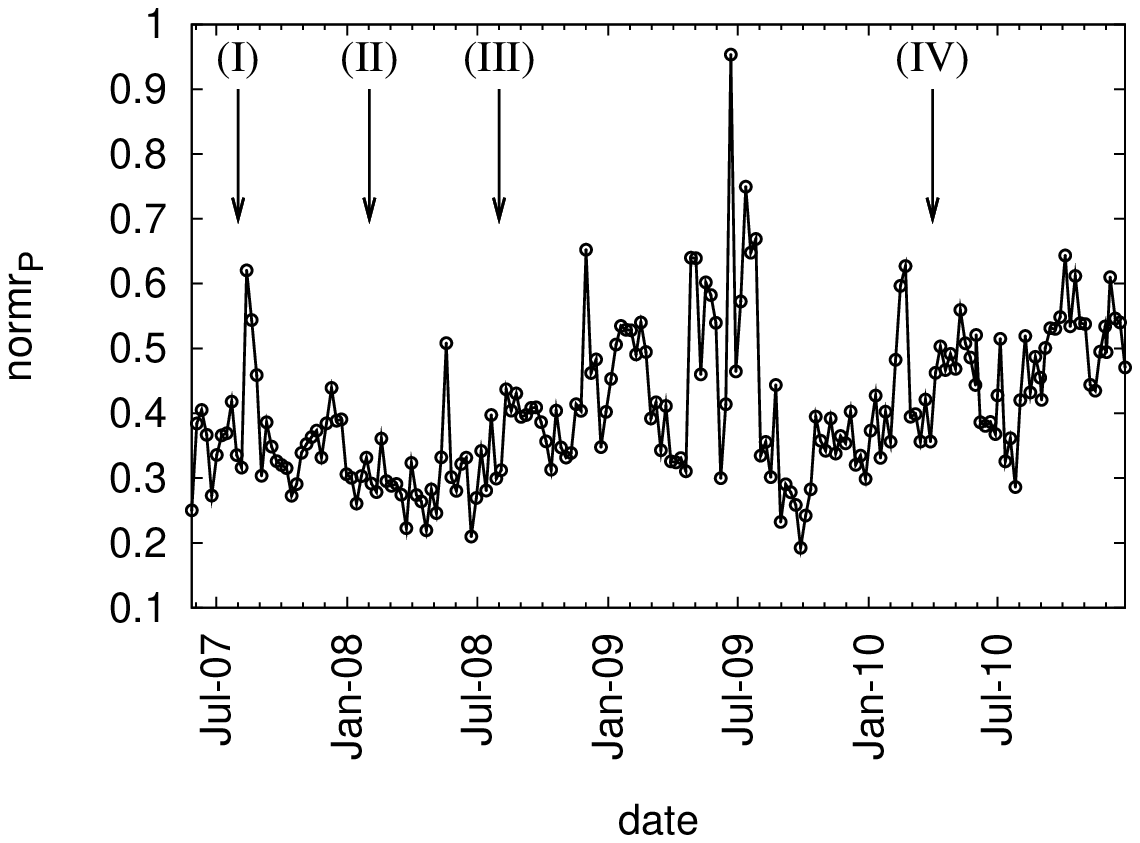}(P)
\includegraphics[scale=0.8]{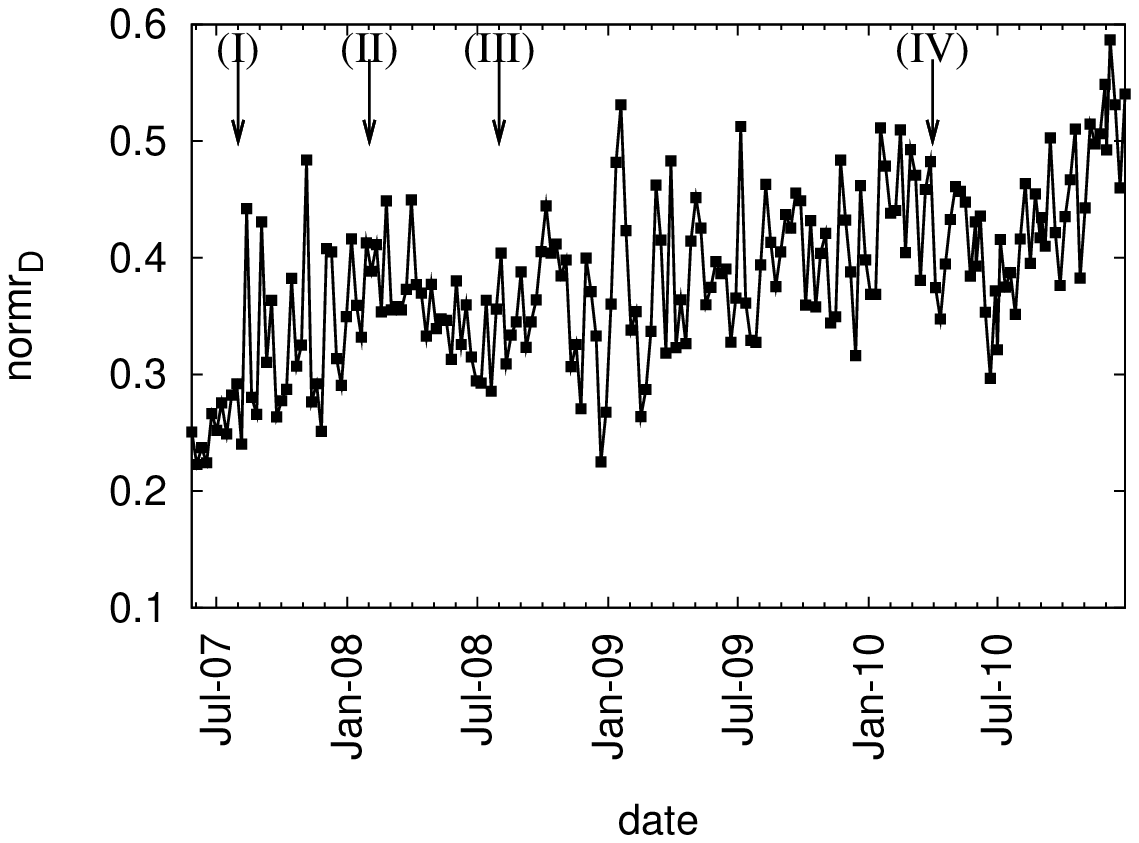}(D)
\caption{The norm of residuals for the
 scaling relationship between the mean of the number of quotations (P) or
 transactions (D) and their standard deviation during a period from June
 2007 to December 2010.}  
\label{fig:FS-residual}
\end{figure}

We show the norms of residuals $normr_{P}$ and
$normr_{D}$ for the scaling relationship for each week in
Fig. \ref{fig:FS-residual}. The figures show that the
statistical quality of the scaling relationship between the mean values
of activities and corresponding standard deviation, the error between the
empirical relationships and Eq. (\ref{eq:scaling2}) with estimates,
depends on the observation period. 

Let us stress that if the norm of residuals is high then mean activities and
their standard deviations may not be fitted to the scaling relation
Eq. (\ref{eq:scaling2}) very well. Specifically, the value of residuals
calculated from the number of quotations suddenly increased or decreased
before or after the periods of the (I) Paribas shock (Aug. 2007), (II)
Bear Stearns shock (Feb. 2008), (III) sub-prime crisis driven by Lehman shock
(Sep. 2008 to Mar. 2009) and (IV) Euro crisis (Apr. to May 2010). The
scaling relationship seems to break after the (I) Paribas shock but before
(III) the sub-prime crisis and (IV) the Euro crisis. The behaviour of residuals
computed from the transaction activities is similar. However, the
scaling relationship of the transaction activities is more stable than
that of the quotation activities before or after the shocks.  

In addition, we compute the global average of the cross-correlation
coefficients. The bootstrap standard deviations are approximately 0.1
during all observation periods. We found that the global average of
simultaneous cross-correlations for quotation activities takes values
less than 0.4 until February 2009. From April to July 2009, it took
larger values than before. After July 2009, the value increased and
fluctuated around 0.5, while the global average of transaction
activities took smaller values. We conclude that the
transactions occur in a mutually independent manner.

\begin{figure}[hbt]
\centering
\includegraphics[scale=0.8]{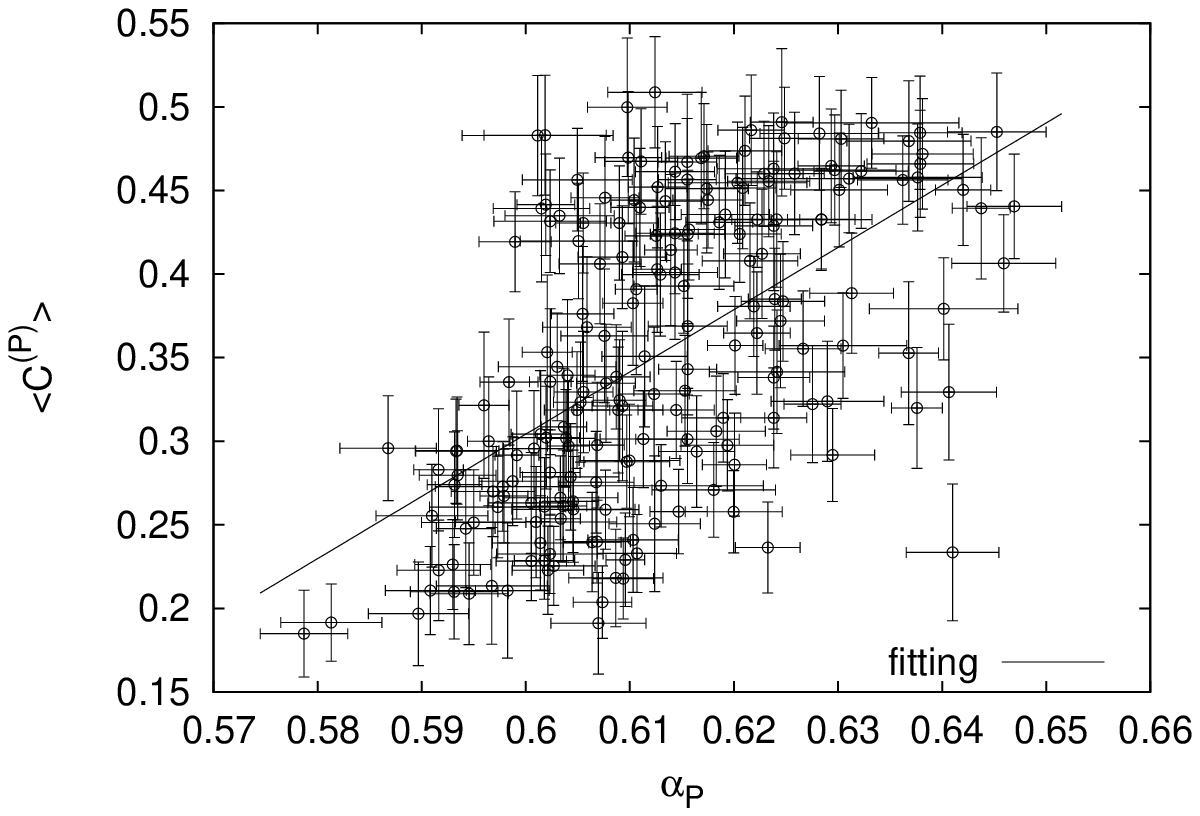}(P)
\includegraphics[scale=0.8]{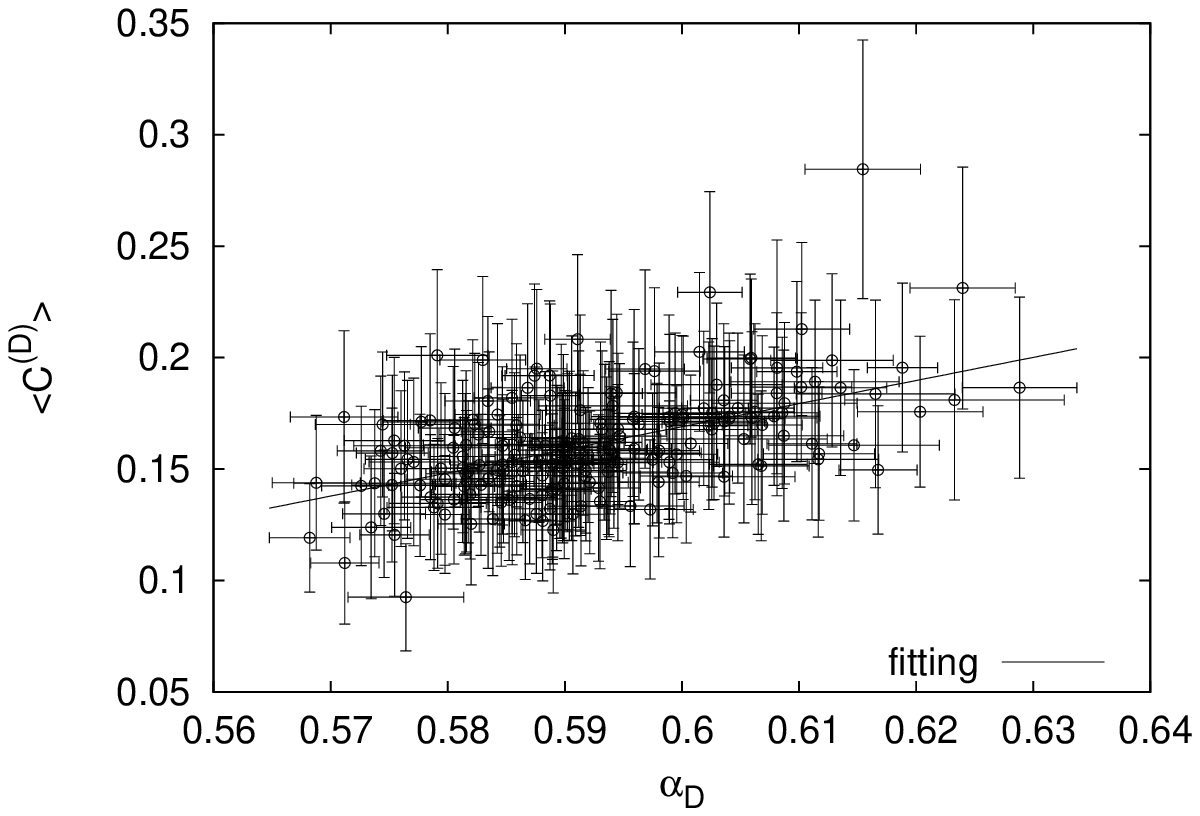}(D)
\caption{Scatter plots between the scaling exponent and
 global average of simultaneous cross-correlation matrix at the same
 observation period. (P) is obtained from the number of quotations, and (D) of
 transactions. The values of the cross-correlation coefficient
 between $\alpha$ and $\langle C_{\Delta t} \rangle$ are 0.464 (P) and
 0.53 (D), respectively. The line represents $\langle C_{\Delta t}^{(X)}
 \rangle = a_X \alpha_X + b_X \quad (X=\{P,D\})$.}
\label{fig:alpha-corr}
\end{figure}

Figure \ref{fig:alpha-corr} shows scatter plots of the scaling exponent
and the global average of simultaneous cross-correlation matrix. Error
bars show a 66.6\% confidence level. The value of the correlation coefficient
between $\alpha_P$ and $\langle C_{\Delta t}^{(P)} \rangle$ is 0.464 for
quotation activities and that between $\alpha_D$ and $\langle C_{\Delta
t}^{(D)} \rangle$ is 0.53 for transaction activities. There is both a
relationship between $\alpha_P$ and $\langle 
C_{\Delta t}^{(P)} \rangle$ and between $\alpha_D$ and $\langle
C_{\Delta t}^{(D)} \rangle$. A solid line represents a fitted linear
dependence $\langle C_{\Delta t}^{(X)} \rangle = a_X \alpha_X + b_X$
obtained with the least squared method, where $X=\{P,D\}$. We found that
$a_P = 3.7231$, $b_P = -1.929$ for quotations and $a_D = 1.0368$,
$b_D = -0.4531$ for transactions. The rms of residuals,
defined as
\begin{equation}
rms_{X} = \sqrt{\sum_{w}\Bigl(\langle C_{\Delta t}^{(X)} \rangle(w) - a_X
 \alpha_X(w) - b_X\Bigr)^2},
\end{equation}
where $\langle C_{\Delta t}^{(X)}\rangle(w)$ and $\alpha_{X}(w)$
are in the observation week $w$, is 0.0734 for quotations ($X=P$) and
0.0195 ($X=D$) for transactions. It is confirmed that a relationship
between the scaling exponent and the global average of
cross-correlations is statistically significant. 


\begin{figure}
\centering
\includegraphics[scale=0.8]{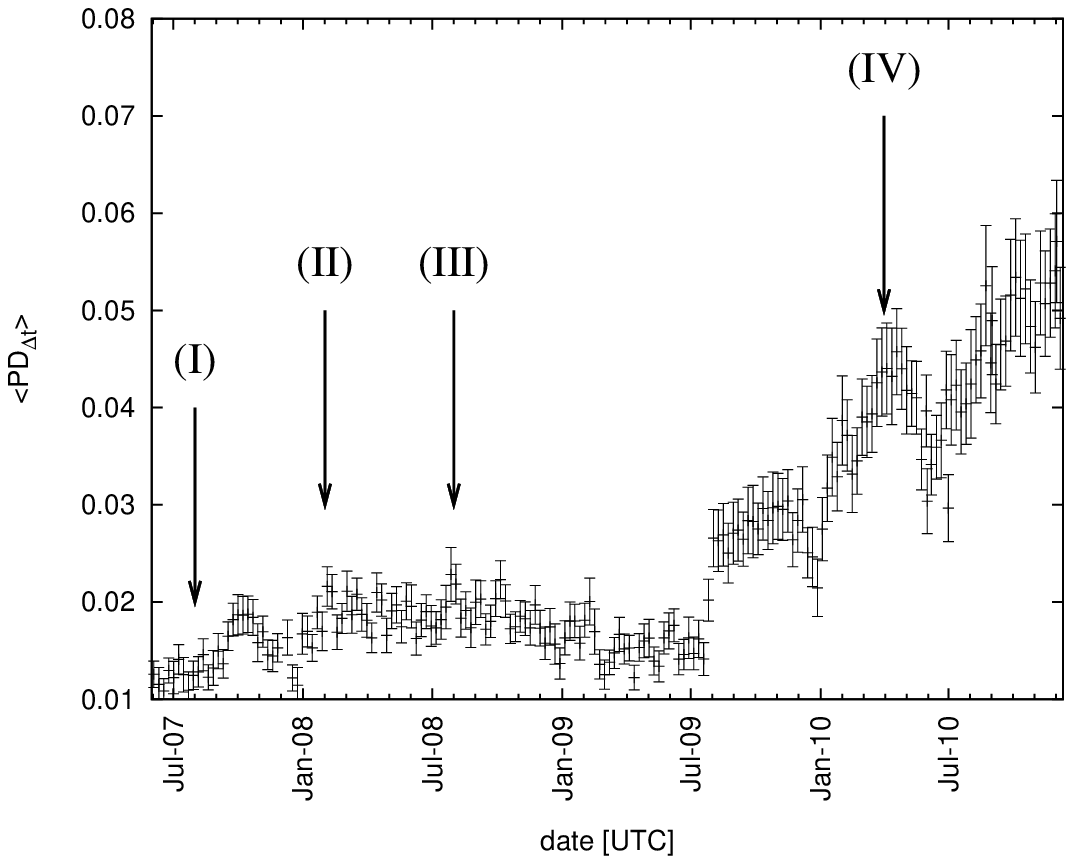}
\caption{The global average of cross-correlations
 between the quotation activities and transaction activities during the
 period from June 2007 to December 2010.}
\label{fig:ave-cross}
\end{figure}

Moreover, using Eq. (\ref{eq:average-pd}), we compute the global average
of cross-correlations between the quotation activities 
and transaction activities during the period from June 2007 to December
2010. Since the global average of their simultaneous cross-correlations
always takes the maximum value, we compute them at
$\tau=0$. Figure \ref{fig:ave-cross} shows their global average for each
week. The error-bars show a 66.6\% confidence level obtained from 
the bootstrap standard deviations. The value of $\langle PD_{\Delta t}
\rangle$ is not large but increases eventually. 
Specifically, after August 2009, the value rapidly increased. This
significant increase coincides with scaling breaking of quotation
activities. From this result, we conclude that in the
EBS brokerage system, the quotation activities changed after August 2009.

\section{Conclusion}
\label{sec:conclusion}
We investigated quotation and transaction activities of the foreign
exchange market with 43 months of high resolution data. We examined
fluctuation scaling, the global averages of cross-correlation coefficients
for both quotation and transaction activities and the global average of
cross-correlation coefficients between them.

We found that the scaling relationship between the mean
of the numbers of activities and their variance is stable for a long
observation period. However, we also observed that the scaling
relationship breaks during several periods. Specifically, the scaling
breaking was observed from May to June 2009. This may be
related to imbalance between demand and supply of the foreign exchange
market. The reason for this is an open question.

We further found that there is a relationship between the
exponents of fluctuation scaling and the global average of
simultaneous cross-correlation coefficients. 
The relationship was found for both different window lengths and different
observation periods. The scaling exponents have a positive association
with the global average of cross-correlations. These values may measure
the degree of the synchronous behaviour of participants.

Moreover, we recognised that the global average of
cross-correlations between the quotation activities and transaction
activities rapidly inclined after August 2009. The
occurred just after the significant scaling breaking of the quotation
activities. The scaling breaking may be related to drastic changes in
participants' perception of the market.

In future research, the relationship between trends of
exchange rates and scaling exponents could be explored. The link between
the intuitive understanding of the prevailing market state by market
participants and estimated scaling exponents should be examined.

\begin{acknowledgement}
This work was supported by the Grant-in-Aid for Young Scientists (B) by
 Japan Society for the Promotion of Science (JSPS) KAKENHI (\#23760074)
 (A.-H. Sato). 
It was also financially supported by the Japan Securities
 Scholarship Foundation, Grants-in-Aid for Scientific Research (A),
 Ministry of Education, Culture Sports, Science and Technology
 (No. 21243019), and by Ishii Memorial Securities Research Promotion
 Foundation (T. Hayashi). This was also supported by the
 European COST Action MP0801 Physics of Competition and Conflicts and by
 the Polish Ministry of Science Grant 578/N-COST/2009/0 (J.A. Holyst). 
\end{acknowledgement}

\end{document}